\newcommand{\sharedpreamble}{%
  \graphicspath{{./}}%
  \usepackage{xurl}%
  \usepackage{hyperref}%
  \definecolor{linkColor}{rgb}{0.8,0,0}%
  \hypersetup{pdfborder={0 0 0},colorlinks=true,linkcolor=linkColor,citecolor=linkColor,urlcolor=linkColor}%
}

\newcommand{\papertitle}{Universal Drift Correction for Multidimensional Scanning Microscopy}

\newcommand{\paperabstract}{
In scanning microscopy, drift causes the specimen to be sampled at positions displaced from the nominal probe positions. 
This displacement alters the spatial assignment of the recorded signals and biases quantitative measurements across two-dimensional imaging, channel-resolved spectroscopic mapping, and scan-position-resolved diffraction analysis. 
Here, we extend orthogonal-scan drift correction from 2D images to spectrum images and diffraction datasets. 
We demonstrate how to recover probe positions using either differently oriented multidimensional scans or structural reference images. 
The recovered positions are used either to resample the multidimensional data onto a regular grid or to assign each recorded signal to its corrected coordinate. 
Our method combines affine and non-rigid correction, requires no prior structural model, and is implemented as open-source, GPU-accelerated software that reduces processing times by two to three orders of magnitude, enabling routine and automated drift correction for quantitative multidimensional microscopy.
}

\newcommand{\paperkeywords}{scan drift; drift correction; spectrum imaging; non-rigid; HAADF; XEDS; EELS; 4DSTEM; STEM; SEM}

\documentclass[11pt]{article}
\usepackage[letterpaper,margin=1in]{geometry}
\usepackage{graphicx}
\usepackage{xcolor}
\usepackage{amsmath}
\usepackage{amssymb}
\usepackage{caption}
\usepackage{microtype}
\usepackage{ragged2e}
\usepackage[numbers,sort&compress]{natbib}
\usepackage{appendix}

\renewenvironment{abstract}{%
  \small
  \par\vspace{0.6em}%
  {\centering\bfseries\abstractname\par}%
  \vspace{0.2em}%
  \noindent\ignorespaces
}{%
  \par\vspace{0.4em}%
}

\makeatletter
\let\@dblfloat\@float
\let\end@dblfloat\end@float
\makeatother

\sharedpreamble
\hypersetup{pdftitle={\papertitle}}

\begin{document}

\RaggedRight

\begin{center}
  {\LARGE\bfseries\papertitle\par}
  \vspace{1.2em}
  {\large
    Sangjoon Lee$^{1,\ast}$,
    William Millsaps$^{1}$,
    Dasol Yoon$^{1}$,
    Caitlyn Obrero$^{1}$,
    Guoliang Hu$^{1}$,
    Corrie Barnes$^{1}$,
    Cedric Lim$^{1}$,
    Andrew Barnum$^{2}$,
    Arthur R.\ C.\ McCray$^{1}$,
    Colin Ophus$^{1,\dagger}$\par}
  \vspace{1.0em}
  {\small
    $^{1}$Department of Materials Science and Engineering, Stanford University, Stanford, CA 94305, USA\\
    $^{2}$Stanford Nano Shared Facilities, Stanford University, Stanford, CA 94305, USA\par}
  \vspace{0.7em}
  {\small
    $^{\ast}$Corresponding author: \href{mailto:bobleesj@stanford.edu}{bobleesj@stanford.edu}\\
    $^{\dagger}$Corresponding author: \href{mailto:cophus@stanford.edu}{cophus@stanford.edu}\par}
\end{center}

\begin{abstract}
\paperabstract

\end{abstract}

\section{Introduction}\label{sec:intro}

Correcting scan drift is a routine processing step across scanning microscopy modalities, including scanning transmission electron microscopy (STEM), scanning electron microscopy (SEM), and atomic force microscopy (AFM) \citep{jin2015,kizu2020,degenhardt2022,diao2023,liu2025}.
Scan drift arises from a combination of specimen, instrument, experimental, and environmental factors.
Temperature changes can cause thermal expansion of the specimen, while specimen charging can produce local displacement or image distortion \citep{li2008full, yothers2017real}.
Instrumental sources include stage creep, mechanical and thermal instabilities of the specimen holder, and instability in the scan hardware \citep{vonharrach1995}.
During \textit{in situ} experiments, temperature ramps and thermal gradients drive continuous thermal drift as the specimen and stage expand, while gas flow, bubbling, and beam-induced reactions cause additional specimen motion or transformation.
Environmental disturbances, including floor and pump vibration together with acoustic noise, can introduce additional scan jitter.

A range of scan-drift correction methods has been developed for two-dimensional scanning microscopy.
Single-image methods estimate scan jitter and linear drift using known periodic crystal structures \citep{jones2013}.
Multi-frame and repeated-scan approaches improve signal-to-noise ratio and enable high-precision strain measurements \citep{berkels2014,yankovich2014,jones2015,jones2017strain,zhang2021optimizing, berkels2019joint}.
Multi-frame and repeated-scan approaches improve signal-to-noise ratio and enable high-precision strain measurements \citep{berkels2014,yankovich2014,jones2015,jones2017strain,zhang2021optimizing}.
Related registration methods align low-signal cryo-STEM image stacks acquired from the same specimen region \citep{savitzky2018}.
Methods based on differing fast-scan directions extend drift correction  without requiring a prior structural model \citep{sang2014,ophus2016drift,ning2018}.
Orthogonal-scan correction has supported quantitative studies of antiferroelectric and toroidal order in oxide superlattices \citep{mundy2022,damodaran2017}, strain fields in twisted bilayer graphene \citep{kazmierczak2021}, short-range order in medium-entropy alloys \citep{zhang2020}, high-entropy thermoelectrics \citep{wang2024}, and grain-boundary vibrations \citep{hoglund2023}.

Advances in detector technology and data acquisition enable scanning microscopy to record multidimensional datasets in which each probe position contains a full spectrum, a diffraction pattern, or both \citep{ophus2019}.
During acquisition, drift causes the probed specimen locations to differ from the nominal scan positions.
The resulting errors affect every signal recorded at those positions and alter structural, chemical, and diffraction-based measurements.

Prior work has established drift correction for multidimensional microscopy through three overlapping groups of methods.
First, measurement-specific methods address particular forms of multidimensional data.
Cross-correlation has aligned energy-filtered image series for chemical and thickness measurements \citep{schaffer2004}.
Non-rigid registration of repeated spectrum-image frames, combined with non-local principal component analysis, has improved elemental maps from low-signal spectrum images \citep{yankovich2016non}.
Nonlinear image correction has also been extended to atomic-resolution energy-dispersive X-ray spectroscopy mapping \citep{kang2025}.
Hyperspectral cathodoluminescence datasets have been drift-corrected using a fast image recorded simultaneously with the spectroscopic acquisition as a drift-free reference \citep{thollar2026}.
Second, structure- and reference-dependent methods require prior structural knowledge or an experimental image treated as undistorted.
Atomically resolved spectrum images and diffraction data have been corrected using prior knowledge of the crystal structure \citep{wang2018}.
Cryogenic 4DSTEM datasets have been remapped to a separately acquired fast STEM image treated as undistorted \citep{smith2023}.
Third, repeated-acquisition methods estimate drift by comparing successive scans.
Multi-pass methods apply this strategy to scan-position-resolved diffraction datasets \citep{mostaed2026}, and non-rigidly registered multi-frame 4DSTEM series have been fused to improve scan-position fidelity and signal-to-noise ratio \citep{oleary2022increasing}.

Here, we present a unified, structure-agnostic drift-correction method for multidimensional scanning microscopy. 
Our method recovers corrected probe positions from two or more scans, with differing fast-scan directions improving correction accuracy.
The corrected positions can be applied to images, spectrum images, and scan-position-resolved diffraction data.
Released as open-source software with GPU acceleration, the method reduces processing time by two to three orders of magnitude relative to the previous implementation, enabling drift and specimen or instrument stability to be assessed during microscope operation.
Because the correction depends on scan geometry and shared structural contrast rather than a modality-specific image-formation model, our method is applicable to a range of raster-scanning modalities. 
We demonstrate it here using atomic-resolution STEM datasets spanning tens of nanometers, including 2D images, XEDS spectrum images, and 4DSTEM datasets.

\begin{figure*}[!t]%
\centering
\includegraphics[width=\textwidth]{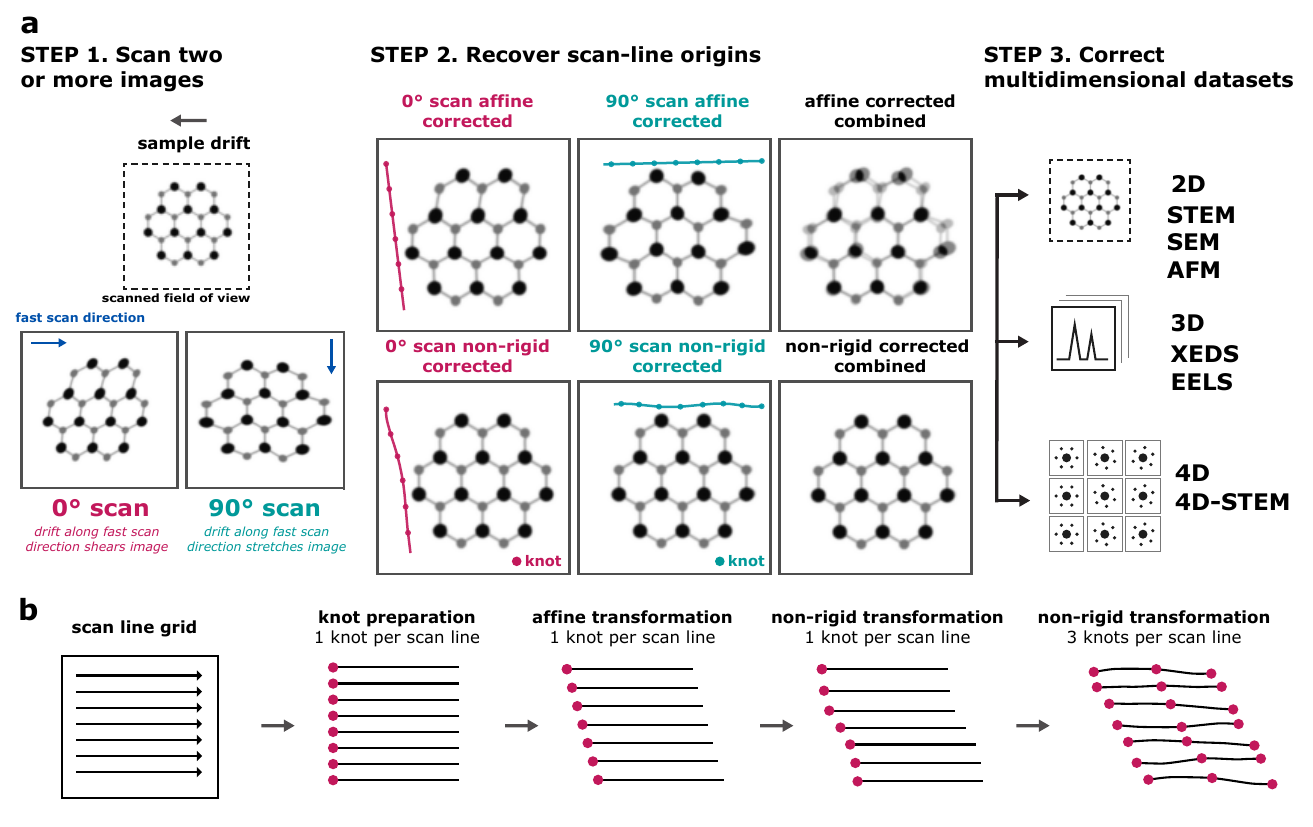}%
\caption{\textbf{Universal drift correction} by recovering corrected scan-line origins for each rastered acquisition. (a) Differing-angle scans record distinct distortions of the same specimen where affine and non-rigid correction stages can recover the probe positions sampled on the specimen. For each scan, the recovered origins are reused for 2D images, channel-resolved multidimensional data, and scan-position-resolved diffraction data. (b) Knot representation of scan geometry, from the initial scan-line grid to affine correction with one knot per scan line and non-rigid correction with three knots per scan line.}\label{fig:overview}
\end{figure*}

\section{Materials and Methods}\label{sec:methods}

\subsection{Overview}\label{subsec:theory}

Our drift correction method estimates linear and non-linear scan drift from two or more scans acquired with different fast-scan directions and recovers the corrected scan-line origins, as outlined in Figure~\ref{fig:overview}(a).
The recovered scan-line origins, combined with each scan's fast-scan direction, define the subpixel probe positions sampled on the specimen.
For 2D imaging, each scan is resampled onto a corrected grid, and the corrected images may then be combined to improve the signal-to-noise ratio.
For 3D spectrum imaging, every energy channel is resampled onto the same corrected grid using the recovered probe positions.
For 4D scan-position-resolved datasets, the two-dimensional signal recorded at each probe position may be resampled across the scan dimensions onto a corrected grid or retained unchanged and associated with the recovered probe positions for downstream analyses.

\subsection{2D affine correction}\label{subsec:affine}

The affine stage first estimates the dominant global linear drift.
Each fast-scan line is treated as rigid, and its origin is represented by a control point, or ``knot'' (Figure \ref{fig:overview} depicts knots as points on a line).
A constant drift rate produces a linear displacement of the scan-line origins with acquisition order.

\begin{figure*}[!t]%
\centering
\includegraphics[width=\textwidth]{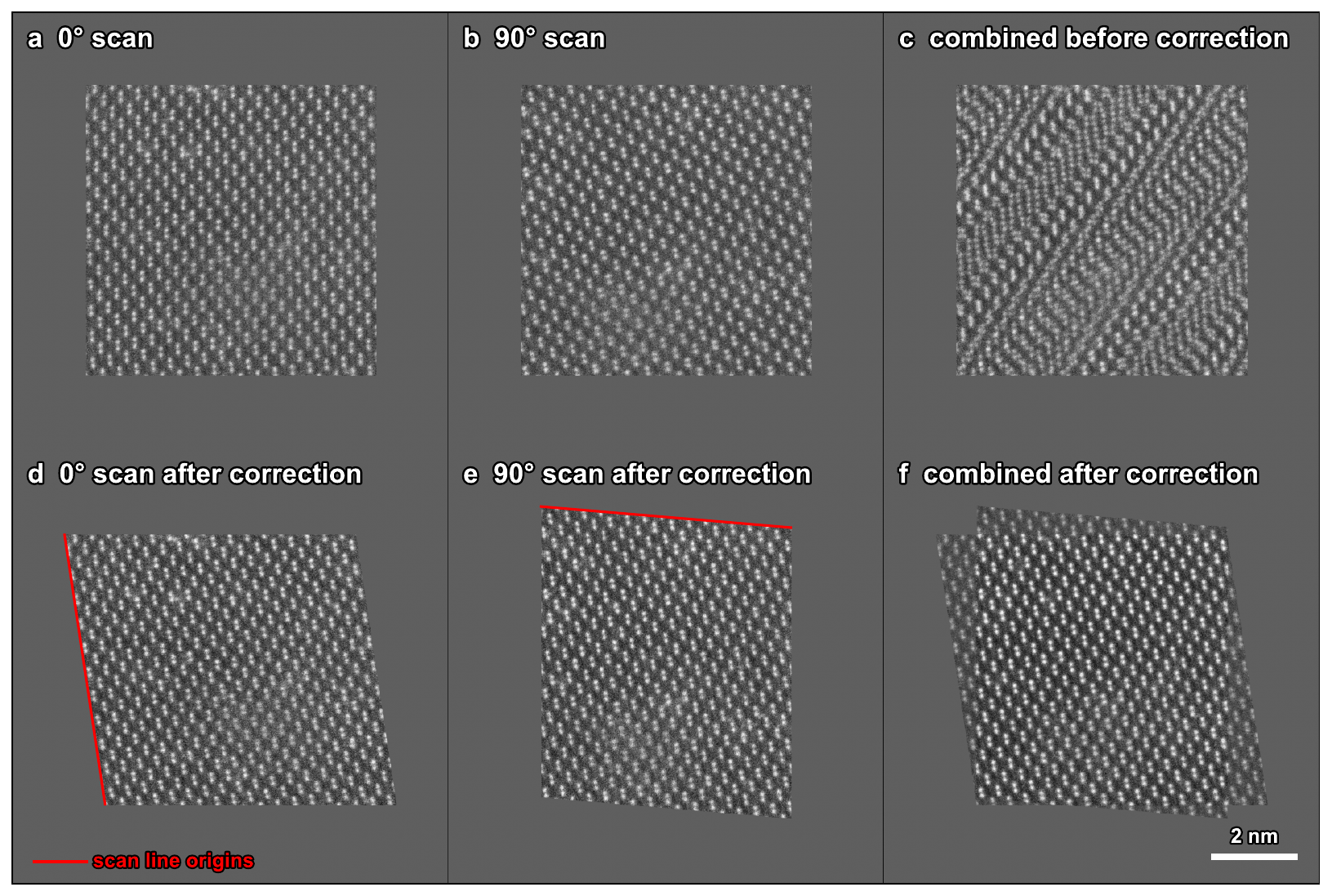}%
\caption{
\textbf{Affine drift correction of a high-angle annular dark-field (HAADF) image} of silicon, using a $0^\circ$/$90^\circ$ scan pair collected from the same specimen region:
(a) $0^\circ$ scan;
(b) $90^\circ$ scan; 
(c) the combined scan before correction;
(d) the $0^\circ$ scan after correction; 
(e) the $90^\circ$ scan after correction, with red lines in (d, e) marking the recovered scan-line origins; and 
(f) the combined scan after only affine correction.
}\label{fig:2d}
\end{figure*}

For each candidate drift, the scans are mapped onto a common output grid and resampled by bilinear interpolation.
A rigid translation is estimated by FFT-based cross-correlation with subpixel peak refinement, and the aligned scans are scored using the metric specified for each dataset.
Image edges may be softened before scoring.
A coarse-to-fine grid search identifies the highest-scoring candidate.
The estimated drift is removed from the knot positions, yielding the affine-corrected scan-line origins.
A final rigid translation aligns the scans, and the translations are mean-centered across the scan set.
The affine-corrected origins then initialize the non-rigid stage.
Panels d and e of Figure~\ref{fig:2d} show the recovered scan-line origins for the orthogonal silicon scan pair. The affine step alone can be enough to correct an image if drift is linear; any significant non-linear drift will require the non-rigid correction.

\subsection{2D non-rigid correction}\label{subsec:nonrigid}

Residual non-linear distortion may remain after affine correction when the drift changes in magnitude or direction during acquisition.
The non-rigid stage therefore allows the affine-corrected fast-scan lines to move independently.
The fast-scan line positions are refined through alternating optimization. 
The standard non-rigid stage uses the Adam optimizer \cite{kingma2014}, while the center-out implementation uses L-BFGS-B. 
The implementation supports normalized cross-correlation and mean-squared-error objectives, with gradient-magnitude images available to reduce sensitivity to contrast differences between scans. 
The optimizer and objective used for each dataset are specified below.
For each image, the non-rigid step optimizes each fast-scan line position by comparing to a reference image. To form the reference image, all other images collected (at least one at a unique scan angle is required) are warped and interpolated with current knot positions and then averaged.
The algorithm cycles through the scans, rebuilding the reference after each update, iterating until the alignment no longer improves.
Regularization steps can stabilize the independent knot updates.
Limits on per-iteration displacement prevent large or unphysical knot movements in regions with weak or repetitive contrast.
Smoothing along the slow-scan direction suppresses high-frequency line-to-line jitter while preserving the overall drift trajectory.
Only a fraction of each proposed update is applied, with a default value of 80\%, to reduce overshooting and promote stable convergence.
For datasets with large drift, the fast scan line positions can be solved by moving outwards from the center of the image. 
A weighted moving average leverages the shifts of previous scan lines to prevent unit cell hops. This is shown in Figure~\ref{fig:2d-nonrigid-center-out}.

\begin{figure*}[!t]%
\centering
\includegraphics[width=\textwidth]{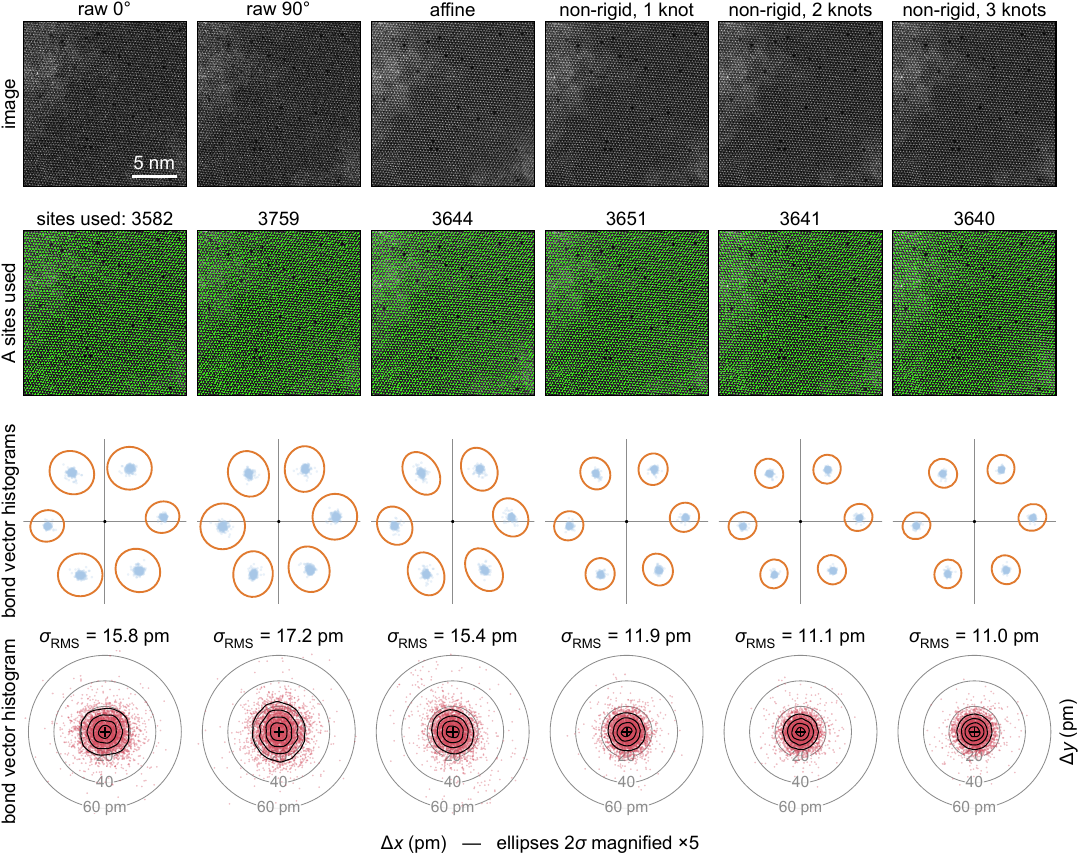}%
\caption{
\textbf{Affine and nonrigid drift correction of WS$_2$ images.}
Columns, from left to right: raw $0^{\circ}$ image, raw $90^{\circ}$ image, affine-corrected image, nonrigid-corrected image assuming no drift within each fast-scan line, and nonrigid-corrected images using two and three control points per fast-scan line. 
Columns 3--6 combine the two raw images, improving the signal-to-noise ratio.
Rows, from top to bottom: processed image, identified A sites, nearest-neighbor bond-vector distributions, and collapsed bond-vector distributions after centroid subtraction.
}\label{fig:2d-nonrigid-knots}
\end{figure*}

\subsection{2D reference-based strip alignment}\label{subsec:strip}

When a drift-corrected reference image or known ground truth is available, a single acquired scan can be aligned to the reference using strip-based correction.
This approach is optional and applies to long-dwell multidimensional acquisitions for which a structural image, such as a high-angle annular dark-field (HAADF) image, is recorded simultaneously with the target dataset.
The first acquired HAADF image is used as the target image and the simultaneously collected HAADF image is divided into strips of consecutive fast-scan lines.
Each strip is aligned independently to the drift-corrected reference using a rigid translation estimated by cross-correlation.
Grouping several scan lines within each strip increases the available signal when individual lines contain insufficient contrast or counts.
The estimated strip translations are interpolated along the slow-scan direction and applied to the corresponding knot positions, producing a smooth set of scan-line origins.
These knot positions can initialize a finer non-rigid refinement when residual line-to-line distortions remain.

\subsection{3D spectrum-image correction}\label{subsec:meth-3d}

In spectrum imaging, a full spectrum is recorded at each scan position. 
Collecting sufficient signal often requires a long total dwell time, making drift a severe issue. 
Two acquisition strategies are commonly used: (1) acquire many fast scans, then register and average them, or (2) acquire a single scan with a long dwell time. 
Approach 1 can minimize drift artifacts in methods such as XEDS. 
For EELS, approach 1 also requires energy alignment to correct zero-loss-peak drift \cite{wang2018towards}; when reliable energy alignment is not feasible, approach 2 may better preserve energy resolution.
To correct drift in approach 2, 
a fast $0^\circ$/$90^\circ$ image pair in the same field of view is acquired and corrected using the affine and non-rigid stages to generate a structural reference.
An image acquired simultaneously with the spectrum image is then aligned to this reference.
This alignment may use affine correction followed by non-rigid refinement, with strip-based alignment available as an optional approach when substantial drift has accumulated.

The next step is to define a regular, integer-spaced output grid for the corrected spectrum image.
Each point on the output grid is mapped back to the corresponding subpixel position in the drifted acquisition.
The full spectrum at the integer position in the output grid is estimated from the neighboring drifted spectra using bilinear interpolation.
The same interpolation weights are applied across all energy channels.

\subsection{4D diffraction-resolved correction}\label{subsec:meth-4dstem}

In diffraction-resolved imaging, a two-dimensional diffraction pattern is recorded at each scan position.
A scalar image suitable for registration is generated from each diffraction-resolved dataset.
This image may be formed by integrating the diffraction intensity over a selected detector region or by applying a phase-retrieval method when phase contrast provides more suitable structural information.
When two or more complete datasets are acquired with differing fast-scan directions, the resulting virtual or phase-retrieved images are aligned using the same affine and non-rigid methods as for two-dimensional images.
The recovered scan-line origins define the probe positions associated with each diffraction pattern.
Alternatively, the diffraction patterns may be resampled along the scan dimensions onto a corrected grid using bilinear interpolation.
This produces corrected virtual images and a resampled 4DSTEM dataset, as shown in Figure~\ref{fig:4dstem}.

\subsection{Multiresolution search and GPU acceleration}
\label{subsec:gpu}

Native-resolution drift searches are computationally expensive for large images, limiting routine correction on local hardware. We therefore implemented batched FFT-based cross-correlation and image resampling on the GPU. For the $2048\times2048$ silicon scan pair shown in Figure~\ref{fig:2d}, the full-resolution implementation required 5.79~s on an NVIDIA RTX PRO 6000 Blackwell GPU, compared with 2440~s for the previous CPU implementation on the same workstation. To reduce the computational requirements further, we introduced a multiresolution strategy that searches for candidate drift rates using a downsampled image, refines the best candidate at progressively higher resolutions, and verifies the final solution at native resolution. This strategy reduced the correction time from 5.79~s to 0.441~s, more than an order of magnitude faster than the full-resolution GPU search, while recovering the drift rates within $6.5\times10^{-4}$ pixels per scan line. The adaptive full-resolution workload enables the workflow more practical on laptop-class and integrated GPUs.

\subsection{Data acquisition}\label{subsec:acq}

All data were acquired on a Thermo Fisher Spectra~300 scanning transmission electron microscope at the Stanford Nano Shared Facilities (nano@stanford).
The microscope was operated at 300~kV with a 30~mrad probe convergence semi-angle and a screen current of 20--150~pA (the WSe$_2$ dataset was acquired at 80~kV).
HAADF and XEDS acquisitions were performed in Velox, with XEDS collected on a Super-X EDS detector.
The 4DSTEM datasets were recorded using a DECTRIS ARINA hybrid-pixel detector \citep{stroppa2023}.

\section{Results}\label{sec:results}

The described method was applied to HAADF images, an XEDS spectrum image, and a 4DSTEM pair.

\subsection{Strong non-rigid drift in \texorpdfstring{WS$_2$}{WS2} and \texorpdfstring{WSe$_2$}{WSe2}}\label{subsec:res-nonrigid}

The WS$_2$ scan pair contained strong drift that varied both within and between fast-scan lines, shown in the first two columns of Figure~\ref{fig:2d-nonrigid-knots}.
The full $2048\times2048$ HAADF images were processed at their native sampling without cropping or a specimen mask.
Each image was independently scaled to $[0,1]$ and padded by 25\% to a $2560\times2560$-pixel canvas using its median intensity.
The two images were warped onto the solver canvas, where pixel intensities and interpolation weights were bilinearly accumulated and then Gaussian-smoothed with $\sigma=0.5$ pixels.
The affine drift rate was found by searching an $11\times11$ grid of row and column rates spaced 0.015 pixels per scan line, then refined on a second pass at 0.003 pixels per scan line, minimizing the mean absolute error between the two frames after low-pass-filtered translation alignment with a softened edge. The grid size and refinement reduction are user-specified.

The affine correction removed the dominant linear drift but left spatially varying misregistration across the field of view.
The WS$_2$ dataset therefore required non-rigid refinement, tested with 1, 2, and 3 knots.
The knots were optimized using Adam for 128 refinement cycles with 30 optimizer steps per cycle.
The mean-squared-error objective used a learning rate of 0.1 and a 32-pixel translation-search bound.
Residual knot trajectories were smoothed along the slow-scan direction with $\sigma=8$ scan lines while preserving their first-order trend.
Each cycle retained 80\% of the proposed knot update. The last two rows of Figure~\ref{fig:2d-nonrigid-knots} demonstrate that the bond-vector distribution of this dataset is sharper with more non-rigid knots.

\begin{figure*}[!t]%
\centering
\includegraphics[width=\textwidth]{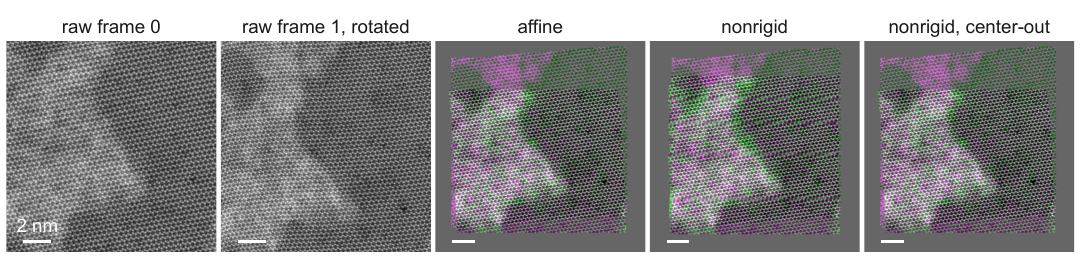}%
\caption{
\textbf{Center-out non-rigid correction in $\bf{WSe_2}$.}
The raw frame exhibits extreme drift that is not fully corrected by standard affine and non-rigid refinement. 
The center-out procedure prevents unit-cell skips.
}\label{fig:2d-nonrigid-center-out}
\end{figure*}

The raw WSe$_2$ scan pair is severely distorted by drift, shown in the first two panels of Figure~\ref{fig:2d-nonrigid-center-out}.
The $4096\times4096$ HAADF images were downsampled to $1024\times1024$ using Fourier cropping, smoothed with a $\sigma=1$ Gaussian, maximum normalized, mean subtracted, and padded by $25\%$. 
Translation by cross-correlation used a 32 pixel taper mask and a Gaussian filter with $\sigma=32$ pixels. Affine correction also used a 32 pixel taper mask and searched a radius cropped $11\times11$ grid (97 points) with candidates spaced 0.015 pixels per scan line, and refined with a second pass of 97 candidates spaced 0.0015 pixels per scan line. Mean absolute error was minimized. A low pass Gaussian filter with $\sigma=32$ pixels was used only for the translation vector; the final affine score used the unfiltered images.
The nonrigid alignment used the center-out algorithm, which uses the shifts of the scan lines closer to the center as a starting point for the slow axis knot position. In the slow axis direction, an exponential moving average with a relative weight of 0.1 applied to the nearest neighbor gives the predicted position. This predicted position is clipped so that the step with respect to the nearest neighbor is within $\pm0.1$ pixel of the global mean step. In the fast axis direction, an exponential moving average with a relative weight of 0.4 applied to the nearest neighbor gives the predicted position. This is seeded to the L-BFGS-B solver with a bounding box of $\pm8$ pixels and a maximum iteration count of 4. The final predicted knot shift is relaxed by 25\%. After each sweep through the knots, a smoothing of $\sigma=16$ was applied to the line of knots. 8 sweeps were used. These parameters are user-selected. The result of the affine, traditional nonrigid, and center-out nonrigid alignments are shown in the last three panels of Figure~\ref{fig:2d-nonrigid-center-out}.

\subsection{2D HAADF imaging across sample types}\label{subsec:res-correction}

\begin{figure*}[!t]%
\centering
\includegraphics[width=\textwidth]{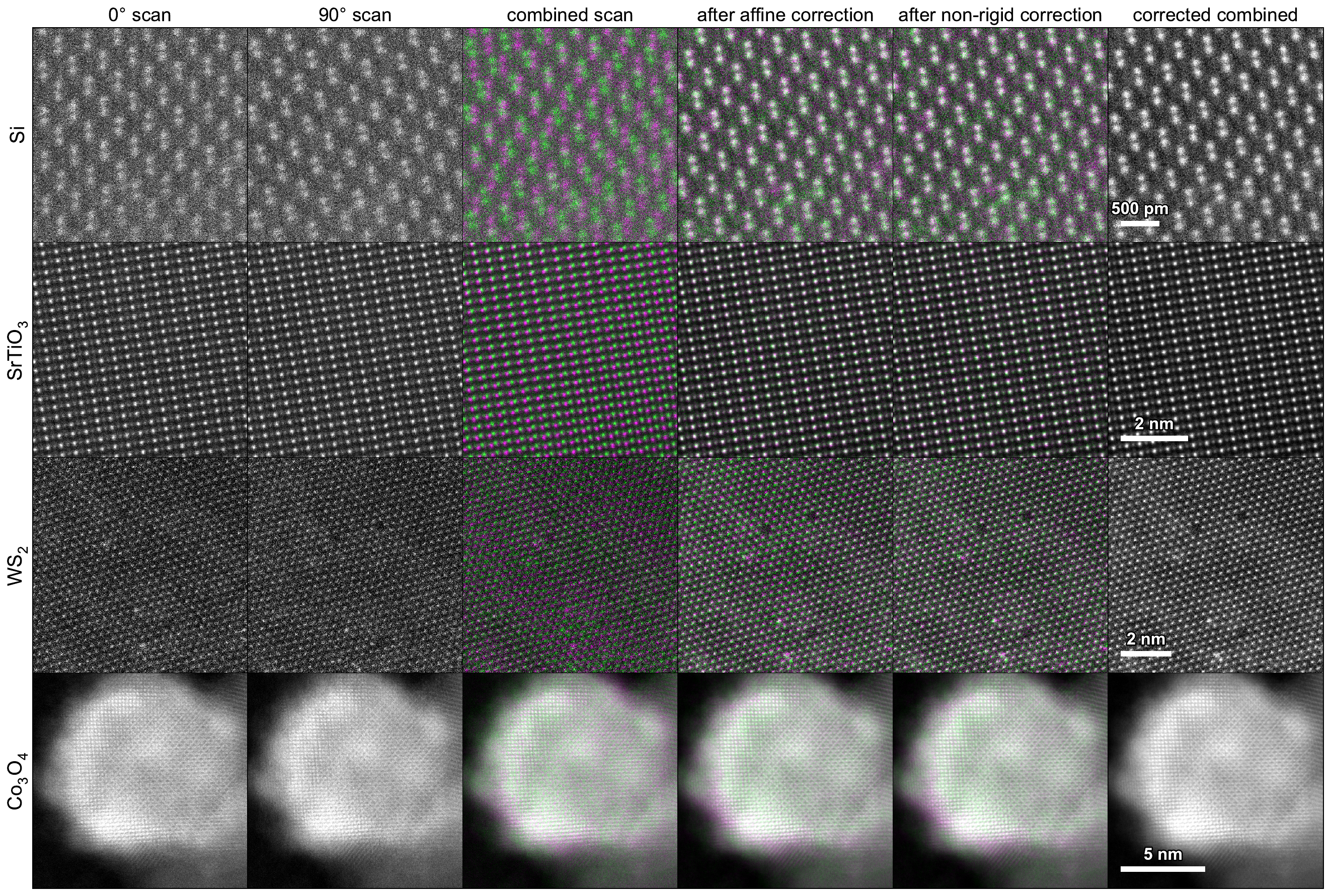}%
\caption{\textbf{Drift correction across sample types.} Rows: silicon, SrTiO$_3$, a WS$_2$ monolayer, and Co$_3$O$_4$. Columns: the $0^\circ$ and $90^\circ$ scans; magenta/green RGB combined views of the two scans before correction, after affine correction, and after non-rigid correction, where misregistration appears as color separation and alignment as white; and the final drift-corrected combined image. All panels of a row share one orientation and field of view, and every panel is rendered by the same pipeline, so the columns differ only by the correction stage.}\label{fig:2d-multiple}
\end{figure*}

For the $2048\times2048$ silicon scans in Figure~\ref{fig:2d},
acquired over a $6.74$\,nm field of view, the correction recovered
a dominant drift component of approximately 284 pixels ($0.93$\,nm).
The same method was applied to $0^\circ$/$90^\circ$ scan pairs of silicon, SrTiO$_3$, WS$_2$, and Co$_3$O$_4$ (Figure~\ref{fig:2d-multiple}).
The WS$_2$ and Co$_3$O$_4$ datasets were also acquired at $2048\times2048$ pixels. The WSe$_2$ dataset was acquired at  $4096\times4096$ pixels.
Across both periodic lattices and aperiodic particle contrast, the corrected orthogonal scans converged to a single set of structural features.
The SrTiO$_3$ dataset exhibited less drift than silicon, but combining the corrected scans still improved the signal-to-noise ratio of the atomic-resolution image.

\subsection{3D XEDS spectrum imaging}\label{subsec:res-eds}

\begin{figure*}[!t]%
\centering
\includegraphics[width=\textwidth]{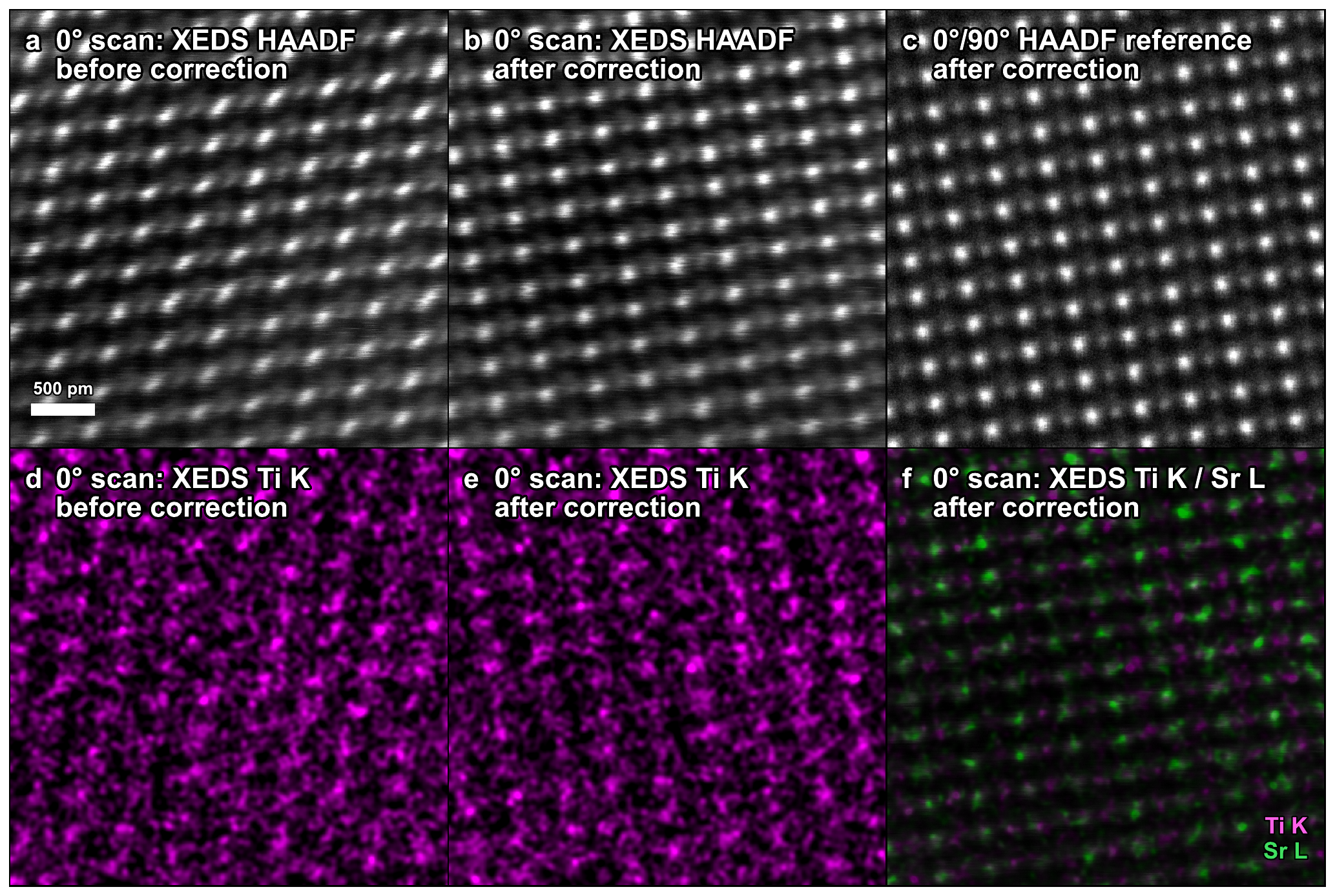}%
\caption{\textbf{Drift correction of a SrTiO$_3$ energy-dispersive X-ray spectroscopy (XEDS) spectrum image.} (a, b) The $0^\circ$ XEDS HAADF before and after correction; (c) the corrected $0^\circ$/$90^\circ$ HAADF reference. (d, e) The Ti~K map (integrated over $4.30$--$4.80$\,keV) before and after correction; (f) the Ti~K (magenta) and Sr~L ($1.70$--$1.95$\,keV, green) composite on the corrected HAADF.}\label{fig:eds}
\end{figure*}

A $2048\times2048$ XEDS spectrum image of a SrTiO$_3$ lattice was acquired over a $6.74$\,nm field of view (Figure~\ref{fig:eds}).
A fast $0^\circ$/$90^\circ$ HAADF pair acquired from the same region was affine-corrected to form a fixed structural reference.
The HAADF images were processed at their native sampling without cropping or a manual specimen mask and were padded by 25\% to $2560\times2560$ pixels using their median intensities.
The $0^\circ$/$90^\circ$ pair was corrected at its native intensity scale, whereas the resulting fixed reference and the simultaneously acquired XEDS HAADF image were each scaled to $[0,1]$ before alignment.
Warped intensities and interpolation weights were bilinearly accumulated on the solver canvas and Gaussian-smoothed with $\sigma=0.5$ pixels.
The automatic multiresolution affine searches evaluated 105 candidates for the reference pair and 265 candidates for the XEDS HAADF image.

The affine result was refined by three sequential strip-correction passes, each registering horizontal bands of consecutive scan lines to the fixed reference by cross-correlation and interpolating the measured translations along the slow-scan direction.
From coarse to fine, the passes used 24, 24, and 64 strips, $(\mathrm{row},\mathrm{column})$ translation limits of $(\pm8,\pm80)$, $(\pm3,\pm12)$, and $(\pm2,\pm6)$ pixels, and smoothing widths of $\sigma=12$, 12, and 6 scan lines.
The first two passes applied the complete measured update, whereas the final pass applied 80\%.
We used a fixed common-coverage mask containing the 75.7\% of pixels sampled by both the structural reference and the final corrected XEDS HAADF image at every correction stage.
Within this mask, the normalized cross-correlation increased from $-0.012$ before correction to 0.693 after affine correction and 0.854 after strip correction.
We apply the same spatial interpolation weights independently to every energy channel, without interpolation along the energy axis.
The Ti~K and Sr~L maps were integrated over $4.30$--$4.80$\,keV and $1.70$--$1.95$\,keV, respectively, and warped using the recovered XEDS scan-line origins.

\subsection{4DSTEM with two orthogonal scans}\label{subsec:res-4dstem}

\begin{figure*}[!t]%
\centering
\includegraphics[width=\textwidth]{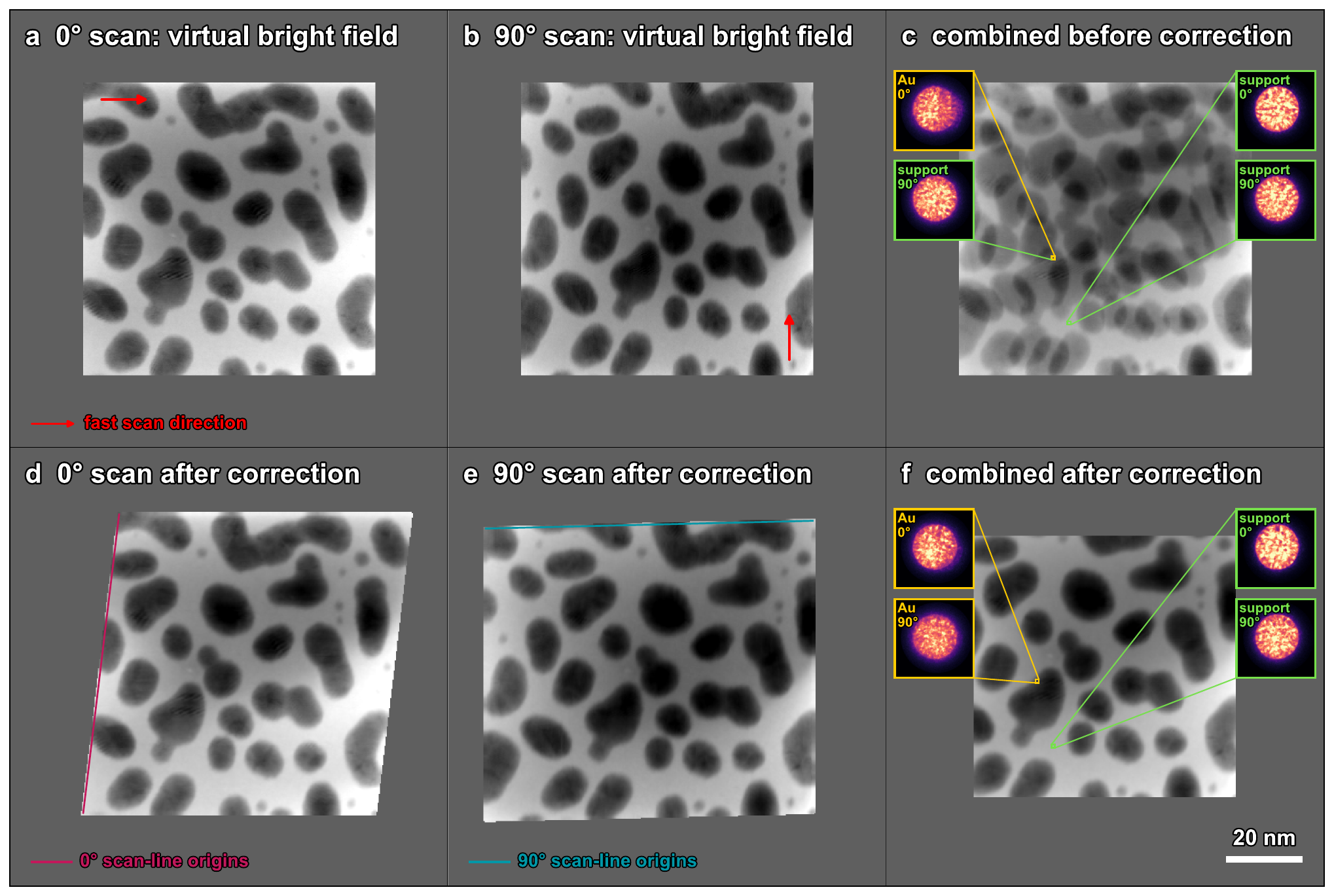}%
\caption{\textbf{Four-dimensional (4D) scan-drift correction of gold nanoparticles.}
(a, b) Virtual bright-field images from the $0^\circ$ and $90^\circ$ acquisitions, with (b) rotated into the $0^\circ$ specimen frame.
(c) Uncorrected combined image.
(d, e) Corrected virtual bright-field images from the two acquisitions.
(f) Corrected combined image.
Insets in (c) and (f) show the individual $0^\circ$ and $90^\circ$ diffraction patterns nearest the selected positions, without spatial averaging or detector-pixel interpolation.
The virtual bright-field images and diffraction patterns use shared display ranges.}\label{fig:4dstem}
\end{figure*}

A complete $0^\circ$/$90^\circ$ pair of 4DSTEM datasets of gold nanoparticles was acquired using the DECTRIS ARINA detector.
The scan-line origins were recovered from scalar virtual images generated from the two acquisitions.
Before correction, the virtual images were misaligned due to scan drift.
After correction, the two virtual bright-field images were combined as shown in Figure~\ref{fig:4dstem}(f).
The recovered probe positions were reused to generate the corrected bright-field images shown in Figure~\ref{fig:4dstem}(d, e).

\section{Discussion}\label{sec:discussion}

Scan-drift correction is well established for two-dimensional scanning microscopy images, but multidimensional datasets also require correction of the probe positions associated with spectra and diffraction patterns.
This method converts the recovered scan-line origins into corrected probe positions for multidimensional datasets.
Once the corrected probe positions are determined, the same geometry can be reused across all signals recorded at those probe positions.
The shared scan geometry provides a unified basis for correcting two-dimensional images, spectrum images, and diffraction-resolved datasets.
The corrected scan geometry is recovered through complementary affine and non-rigid stages.

Both stages operate on the same knot-based representation of the scan-line origins.
The affine stage constrains the knots to a global linear trajectory that captures the dominant drift, whereas the non-rigid stage allows independent knot motion to describe changes in drift and local line-to-line deviations.
This two-stage design uses a constrained model when it is sufficient while retaining the flexibility required for more complex scan trajectories.
The correction relies on scan geometry and shared structural information, not on a STEM-specific image-formation model. The same formulation can be adapted to other raster-scanning modalities, including SEM and AFM.

For multidimensional datasets, the recovered probe positions define where each recorded signal was sampled on the specimen.
In spectrum imaging, each energy channel is resampled onto the corrected integer grid, preserving the full spectrum at each corrected location for elemental mapping, spectral fitting, and chemical quantification.
For diffraction-resolved datasets, the diffraction patterns may remain unchanged while the associated probe positions are updated.
The updated positions can then be supplied directly to strain mapping, orientation mapping, and ptychographic reconstruction.
Providing drift-corrected probe positions in advance may reduce positional uncertainty and the number of position parameters that must be refined during iterative ptychographic reconstruction \citep{maiden2012annealing, mccray2025}.
Alternatively, the diffraction-resolved dataset may be resampled along the scan dimensions when a regular corrected grid is required for visualization.

The correction assumes that corresponding structural features remain identifiable across the input scans or between the target dataset and its reference.
Weak contrast, low signal-to-noise ratio, or repetitive structures may produce ambiguous alignments even when the optimization converges.
In periodic structures, this ambiguity may cause a scan line or knot to converge to a lattice-equivalent position separated by one or more unit cells during non-rigid refinement. 
Solving outward from the image center or a local fiducial, such as a beam-induced hole or distinctive interface, can mitigate this artifact.
Changes in specimen structure or imaging conditions between acquisitions, including beam damage, contamination, defocus variation, and structural evolution, may also reduce agreement between the scans and bias the recovered positions.

When collecting 2D scanned data, we recommend that the user collect a pair of images with orthogonal fast scan directions. 
Any additional images will improve accuracy, but in practice, two images almost always provide sufficient improvement. 
For 3D spectrum data, we recommend single scans, rather than multi-pass scans.
When processing data, we recommend using the translation, affine, and nonrigid alignments. When datasets are challenging, the user may enable edge softening and the center-out nonrigid alignment. For scanning techniques that have Z drift, we recommend plane fitting and subtraction before using this algorithm.
We recommend a visual validation of the merged output image, especially for periodic datasets.

We have also created a GPU-accelerated, open-source Python implementation in PyTorch \citep{paszke2019} to support drift correction during microscope operation.
The recovered scan line origins could also be written back to the scan generator to offset the upcoming raster, as in live predictive drift compensation \citep{mosse2026}.
Integration with microscope and detector-control software could further automate the transfer of the recovered scan geometry to spectrum analysis, diffraction mapping, strain and orientation mapping, ptychographic reconstruction, and \textit{in situ} measurements of structural dynamics \citep{lee2026}.
Together, the reusable probe positions and acquisition-time processing provide a unified route to drift assessment and correction during experiments, supporting more reliable quantitative analysis of multidimensional scanning data.

\section{Conclusion}\label{sec:conc}

We have demonstrated a unified method for correcting scan drift across two-dimensional images, channel-resolved spectrum images, and scan-position-resolved 4DSTEM datasets.
By recovering the probe positions sampled on the specimen, the method supports downstream analyses, including strain and orientation mapping, atomic-scale spectroscopic mapping, phase mapping, and iterative ptychographic reconstruction.
GPU acceleration reduces the correction time by two to three orders of magnitude, allowing the affine and non-rigid stages to complete on acquisition-relevant timescales.
Released as open-source software with curated tutorials and reproducible workflows, the method offers an accessible route for drift correction in multidimensional scanning microscopy.

\begin{appendices}

\end{appendices}

\section{Data Availability}

The experimental datasets, analysis code, and figure-generation scripts supporting this study are available in Dryad (\url{https://doi.org/10.5061/dryad.7d7wm38bs}).

\section{Code Availability}

The drift-correction method is implemented in the open-source \texttt{quantem} Python package (\url{https://github.com/electronmicroscopy/quantem}).
Tutorial notebooks reproducing the workflows are available at \url{https://github.com/electronmicroscopy/quantem-tutorials}.
For questions about implementation or support, readers may contact the corresponding authors by email.

\section{Competing interests}

The authors declare no competing interests.

\section{Author contributions statement}

CO implemented the original orthogonal-scan non-rigid drift-correction in Python and supervised the work.
SL implemented the reference strip-based alignment algorithm, multi-dimensional dataset correction, and GPU acceleration, and performed analysis of the results.
WM implemented the center-out algorithm and performed non-rigid analysis on WS$_2$.
SL, WM, DY, and CO wrote the manuscript.
DY, WM, SL, COb, GH, CB, and CL acquired microscopy data.
AB and CL developed automated scanning acquisition at the Stanford Nano Shared Facilities (nano@stanford).
ARCM implemented the algorithm for applying drift adjusted scan positions for iterative ptychographic reconstruction.
All authors participated in reviewing, editing, and approving the final text.

\section{Acknowledgments}

We thank Samsung Electronics Corporation for providing the silicon samples and Kevin Crust and his advisor, Harold Y. Hwang, for providing the strontium titanate sample. 
We thank Jun Beom Hwang and his advisor, Myoung Hwan Oh, for preparing the cobalt oxide sample. 
We thank Pinaki Mukherjee at the Stanford Nano Shared Facilities (nano.stanford.edu) for TEM instrument support and Berk Kucukoglu and Oliver Harder (DECTRIS) for supporting 4DSTEM ARINA data acquisition and automated acquisition workflows. 
We thank Denis Leshchev (NVIDIA) for 4DSTEM data acquisition and streaming support.
We also thank Amy McKeown-Green and Stephanie Ribet for providing experimental data for testing. Authors thank Lauren Hoang and Anh Tuan Hoang for preparing the WS$_2$ and WSe$_2$ samples, and their advisors Andrew Mannix and Eric Pop for their support.

\section{Funding sources}

This research was supported by the U.S. Department of Energy under Contract No. DE-AC02-76SF00515 through the Basic Energy
Sciences (BES) Microelectronics ESTEEM Program
(101256). 
We also acknowledge additional support from Samsung Electronics Co. Ltd. (IO250812-13405-01), the Toyota Research Institute, and Stanford University.

\nocite{*}
\bibliographystyle{unsrtnat}
\bibliography{reference}
\end{document}